\documentclass[aps,showpacs,twocolumn,amsmath,amssymb,amsfonts,eqsecnum]{revtex4-1}
\usepackage{graphicx}
\usepackage{color}
\usepackage{natbib}
\usepackage{epstopdf}
\usepackage{hyperref}
\usepackage{multirow}
\usepackage{ulem}

\begin{document}
\title{Reply to {\it Comment on "Casimir force in the $O(n\to\infty)$ model with free boundary conditions"}}
\author{Daniel Dantchev$^{1,2}$\thanks{e-mail:
daniel@imbm.bas.bg}, Jonathan Bergknoff$^{1}$\thanks{e-mail: jbergk@physics.ucla.edu} and  Joseph Rudnick$^{1}$\thanks{e-mail:
jrudnick@physics.ucla.edu}} \affiliation{ $^1$ Department of Physics and Astronomy, UCLA, Los Angeles,
California 90095-1547, USA,\\$^2$Institute of
Mechanics - BAS, Academic Georgy Bonchev St. building 4,
1113 Sofia, Bulgaria
}
\date{\today}
\begin{abstract}
The proceeding comment  raises a few points concerning our paper Dantchev \textit{et al.}, Phys. Rev. E. {\bf 89}, 042116 (2014). In this reply we stress that while Refs. Diehl \textit{et al.} EPL {\bf 100}, 10004 (2012) and Phys. Rev. E. {\bf 89},
062123 (2014)  use three different models to study the  the Casimir force for the $O(n \rightarrow \infty)$ model with free boundary conditions we study a single model over the entire range of temperatures, from above the bulk critical temperature,  $T_c$, to absolute temperatures down to $T=0$. The use of a single model renders more transparent the crossover from effects dominated by critical fluctuations in the vicinity of the bulk transition temperature to effects controlled by Goldstone modes at low temperatures. Contrary to the assertion in  the comment, we make no claim for the superiority of our model over any of those considered by Diehl \textit{et al}. We also present additional evidence supporting our conclusion in Dantchev \textit{et al.}, Phys. Rev. E. {\bf 89}, 042116 (2014) that the temperature range in which our low-temperature analytical expansion for the Casimir force  increases as $L$ grows and remains accurate for values of the ratio $T/T_c$ that become closer and closer to unity, while  $T$ remains well outside of the critical region. 
\end{abstract}

\maketitle

Before responding to the points raised in the comment by Diehl \textit{et al.} \cite{DGHHRS2014c} we summarize our article \cite{DBR2014}. Making use of the spherical model version of a lattice of fixed length spins  with nearest neighbor interactions, we calculated the thermodynamic Casimir force of a three dimensional cubic lattice of such spins that is infinite in two dimensions and that has a finite extent, $L$, in the third. We applied free boundary conditions in the third dimension, which are more physically relevant than the more tractable, and well explored, option of periodic or anti periodic boundary conditions. Because of this choice it was necessary to implement separate spherical constraints layer by layer in that direction. Our calculations yield numerical results for the Casmir force that span all interesting temperature ranges, from above the bulk critical point down to absolute zero. When $|T-T_c|L$ is not too large---that is, in the finite size scaling regime---our results are consistent with finite size scaling, in that the Casimir force per unit area takes the form $F_{\rm Cas}(T,L)/A = k_BT_c L^{-3} \vartheta((T-T_c)L)$, with $\vartheta$ a universal function of the scaling combination $ x=(T-T_c)L$. All our results in this region agree with those obtained by Diehl and co-workers \cite{DGHHRS2012,DGHHRS2014}. By contrast, outside of that regime, and most particularly at low temperatures, we discovered that the Casimir force is inconsistent with finite size scaling, in that there is a contribution that cannot be incorporated into the scaling form above. 

We supplemented our numerical results with an expansion about the zero temperature limit of the Casimir force. Making use of this expansion we obtained an analytical expression for the Casimir force that is asymptotically exact at $T=0$, that is also exact for the leading order expansion about that limit. Furthermore, the expression agrees with our numerical results to an excellent degree of accuracy over a wide range of temperatures, excluding of course the finite size scaling region. It is important to note that the model we studied was viewed for some time as analytically intractable \cite{BJSW74,BM77}. Nevertheless, we were able to extract analytical results at temperatures well below the critical one, and in addition, provide indications that it may be possible to derive exact results at the critical point. All the steps in our investigation are laid out in the paper so that anyone interested in doing so can reproduce our results, extend or critically evaluate them, or make use of our methods in another context. 

By contrast, looking at the same geometry, Diehl \textit{et al.} utilize the Ginzburg-Landau-Wilson (GLW) effective Hamiltonian, studying two versions of the $n \rightarrow \infty$ $O(n)$ model, as described in the comment and in \cite{DGHHRS2012,DGHHRS2014}. They focus primarily on behavior in the finite size scaling regime, performing exhaustive and sophisticated analyses to extract results in that regime that are accurate to a remarkable number of significant figures. Taking limits in a particular way, they map on to earlier findings of Bray and Moore on the semi-infinite $O(n \rightarrow \infty)$ model with Dirichlet boundary conditions. In the vicinity of the bulk critical point the authors focus on two GLW models, model A in which spins are confined to discrete layers perpendicular to the bounding surfaces but are continuous distributed within each layer and model B in which spins lie on a two dimensional lattice within those layers. Unsurprisingly, corrections to scaling differ between the two models, while those models exhibit identical asymptotic critical behavior.  For low temperature calculations they map onto a non-linear sigma model which for brevity we denote as model C. According to the authors model C, the analysis of which is not detailed, predicts asymptotic low temperature behavior that is consistent with finite size scaling, in that it can be expressed as a function of the scaling variable $x$. 

It is important to note that the three models above belong to the same universality class as the model we consider. It is also important to keep in mind that membership in the same universality class does not guarantee identity of thermodynamic quantities outside of the critical regime. Even though a uniaxial antiferromagnet and a gas-liquid system have the same equilibrium critical exponents at their respective critical points, a liquid freezes at sufficiently low temperatures and high pressures, while the typical antiferromagnet exhibits no corresponding behavior. No evidence has been presented that the model we consider has the same low temperature thermodynamic Casimir force as the nonlinear sigma model, nor has any argument been presented that the nonlinear sigma model utilized in  \cite{DGHHRS2012,DGHHRS2014} will reproduce the low temperature Casimir force of either model A or model B. 

We stress that our study is performed on a single model. We make no claims for the superiority of this model over any of those considered in \cite{DGHHRS2012,DGHHRS2014}. Rather, we emphasize that our use of that one model to explore the Casimir force over the entire range of temperatures renders more transparent the crossover from effects dominated by critical fluctuations in the vicinity of the bulk transition temperature to effects controlled by Goldstone modes at low temperatures.

As the comment  \cite{DGHHRS2014c} makes clear, but which is not evident in the articles \cite{DGHHRS2012,DGHHRS2014},  the low-temperature behavior  studied there corresponds to $T<T_c$  but with $-t\lesssim 1$, where their definition of $t$ is  $t=-4\pi(R-R_{\mathrm{c}})$ with $R=\beta J$, which implies $T/T_c \gtrsim 0.76$, i.e., below but nevertheless still {\it close} to $T_c$.  In other words, the authors of the comment simply did not study ``temperatures considerably below that of the bulk transition", as asserted in \cite{DBR2014}. 

We find that, if one fixes $R \sim 1/T$ large enough, then for any $L$ the Casimir forces for different $L$'s cannot be made to fall on a single curve by invoking finite size scaling. This is in fact clearly shown in the plot displayed in \cite{DGHHRS2014c} as well as in Figure 3 in our article \cite{DBR2014}. Furthermore, we find that the numerically determined Casimir force is well reproduced by our analytical expressions. That is to say, those expressions accurately predict the Casimir for systems with fixed $L$ at low temperatures. We agree with the statement in the comment that our ``expansion never captures the correct asymptotic behavior", i.e., the  the scaling behavior of the Casimir force in the critical regime. On the other hand, the finite size scaling form obtained in \cite{DGHHRS2012,DR2014,DGHHRS2014} does not reproduce the correct {\it true} behavior of the Casimir force at fixed $L$ for low enough temperatures. The absence of finite size scaling in the low temperature regime is by no means self-evident, given that when  periodic or anti-periodic boundary conditions are imposed in the finite direction \cite{D96,D98,DG2009,SR86} finite size scaling holds down to $T=0$. This leads to the infererence that the presence or absence of finite size scaling in the low-temperature, Goldstone mode dominated, regime strongly depends on boundary conditions. On the other hand, for all boundary conditions now studied the limit $\lim_{L\to\infty}\lim_{T\to 0}\lim_{A\to\infty}[L^{3}F_{\rm Cas}(T,L)/( k_B T A)]$ is universal; for  periodic and Dirichlet boundary conditions this limit is given by the value of the scaling function of the force for the Gaussian model at its critical temperature when subject to the those boundary conditions.

By contrast, the results reported in \cite{DGHHRS2012,DGHHRS2014}, including those pertinent to their ``low-temperature regime", are formulated in terms of the scaling variable $x= -4\pi (R-R_c) L$.  We now understand that our original interpretation of that regime as encompassing temperatures close to absolute zero was in error. The error is understandable given that it is natural to assume that a low temperature regime would refer to low absolute temperatures, rather than temperatures for which $T/T_c \gtrsim 0.76$.

With the above in mind we hope it is clear why we stated that we ``partially corrected'' the results of \cite{DGHHRS2012} in the low-temperature regime, while at the same time we are in full agreement with their results for the scaling regime. In fact, we make a special point in \cite{DBR2014} of noting that our low temperature results are inconsistent with finite size scaling, so it is no criticism of those results to point out that they do not apply in the finite size scaling regime. In fact, we consider the inconsistency of our low temperature results with finite size scaling to constitute strong evidence that the Casimir force at low temperatures arises from different physical mechanisms (Goldstone modes) than does the critical Casimir force.

With regard to the assertion in the comment that the results in \cite{DGHHRS2012} and \cite{DGHHRS2014} are ``numerically exact'' we note that  \cite{DGHHRS2014} contains a correction of the low temperature Casimir force as reported in \cite{DGHHRS2012}; see the footnote on page 16 in Ref. \cite{DGHHRS2014}. As for the phrase ``numerically exact,'' the one interpretation that makes sense, as we see it,  is that the authors utilize an algorithm that produces predictions that are as accurate as the implementation of that algorithm allows. By this interpretation our results are also numerically exact.  In any case, we did not dispute the numerical precision of the data presented in \cite{DGHHRS2012} and \cite{DGHHRS2014} for a fixed $L$. We simply pointed out that a related but not identical model analyzed at low temperatures provides information about the behavior of the thermodynamic Casimir force that is not subsumed in the analysis performed by the authors. Furthermore we are able to present exact closed-form analytical results for the Casimir force in the very low temperature regime ($T$ asymptotically close to absolute zero).

Our low temperature results are criticized in the comment, in which the argument is made that our reliance on an expansion in absolute temperature limits the applicability of our expression to very low temperatures. However, it is a numerical  fact that our analytical results quite accurately reproduce  the behavior of the Casimir force for  reasonably large values of $L$ and for all $T<0.8 \;T_c$, as shown in Fig. 1 of our article \cite{DBR2014}. In fact, the larger $L$ is the better we find the approximation to be. We take this to be a strong empirical argument for the utility of the expansion. To further quantify our findings about the applicability of the low temperature expressions, the table below displays that values of $T/T_c$ below which the expression is accurate to both 5\% and 10\%. 
\begin{table}[h!]
\label{table}
\begin{tabular}{|c|c|c|c|c|c|c|c|}
\hline & L & 50 & 100 & 150& 200 & 300 & 500 \\ 
\hline 
\multirow{2}{*}{5 \%} & $T/T_c$ & 0.89 & 0.901 & 0.913 & 0.922 & 0.935 &  0.949 \\ 
& x* & -19.6 & -34.8 & -45.5 & - 53.8 & -66.52 & -84.6 \\
\hline  
\multirow{2}{*}{10 \%} & $T/T_c$ & 0.938 & 0.947 &0.953 & 0.958 & 0.965 & 0.974 \\ 
&x* & -10.5 & -17.8 & -23.4 &  - 27.8 & -34.3 & -41.9\\
\hline 
\end{tabular} 
\end{table}
The bounding ratios are shown for  values of $L$ ranging from 50 to 500. Note that when $L=500$, 5\% accuracy is achieved for all values of $T/T_c$ that are less than about 0.95.  

The  accuracy of our expressions over a broad range of temperatures does lead one to ask why that is so. In \cite{DBR2014} we presented arguments that our analytical expressions should be accurate when $-x\equiv 4\pi (R-R_c)L \gg \log L$. As it turns out, the table above has $x^*$  increasing with $L$. That increase is slower than linear in $L$. The $5\%$ data for $x^*$ are perfectly fitted by
$-21.98 (1+3.06 \log L-8.84 \log \log L)$. In any case, it is clear that as $L$ increases our expressions remain accurate for values of the ratio $T/T_c$ that become closer and closer to unity. While the question of the exact range of applicability of our analytical expression is still open, the actual numbers do support our contention that the range of accuracy of our expression grows with increasing $L$. 

This leads us to the following statement in  \cite{DGHHRS2014c}:  ``our numerical results presented in Fig. 1 show that the range in temperature where the behavior is well
described by the scaling function essentially does not depend on the thickness $L$, or equivalently, the range in the
scaling variable $x$ increases proportional to $L$''. This  is inconsistent with our claim for the expanding accuracy of our low temperature expression. Furthermore it conflicts with the  standard implementation of finite-size scaling theory \cite{FB72,Ba83,P90,BDT2000}, which is based on the expectation that the finite size scaling regime, in which the behavior of the finite system differs essentially from that of the infinite system, shrinks in absolute temperature as $L$ increases. Our results, on the other hand, are in a full conformity with this expectation.


\begin{thebibliography}{15}
\expandafter\ifx\csname natexlab\endcsname\relax\def\natexlab#1{#1}\fi
\expandafter\ifx\csname bibnamefont\endcsname\relax
  \def\bibnamefont#1{#1}\fi
\expandafter\ifx\csname bibfnamefont\endcsname\relax
  \def\bibfnamefont#1{#1}\fi
\expandafter\ifx\csname citenamefont\endcsname\relax
  \def\citenamefont#1{#1}\fi
\expandafter\ifx\csname url\endcsname\relax
  \def\url#1{\texttt{#1}}\fi
\expandafter\ifx\csname urlprefix\endcsname\relax\def\urlprefix{URL }\fi
\providecommand{\bibinfo}[2]{#2}
\providecommand{\eprint}[2][]{\url{#2}}

\bibitem[{\citenamefont{{Diehl}
  et~al.}(2014{\natexlab{a}})\citenamefont{{Diehl}, {Gr{\"u}neberg},
  {Hasenbusch}, {Hucht}, {Rutkevich}, and {Schmidt}}}]{DGHHRS2014c}
\bibinfo{author}{\bibfnamefont{H.~W.} \bibnamefont{{Diehl}}},
  \bibinfo{author}{\bibfnamefont{D.}~\bibnamefont{{Gr{\"u}neberg}}},
  \bibinfo{author}{\bibfnamefont{M.}~\bibnamefont{{Hasenbusch}}},
  \bibinfo{author}{\bibfnamefont{A.}~\bibnamefont{{Hucht}}},
  \bibinfo{author}{\bibfnamefont{S.~B.} \bibnamefont{{Rutkevich}}},
  \bibnamefont{and} \bibinfo{author}{\bibfnamefont{F.~M.}
  \bibnamefont{{Schmidt}}}, \bibinfo{journal}{the preceding comment}.

\bibitem[{\citenamefont{Dantchev et~al.}(2014)\citenamefont{Dantchev,
  Bergknoff, and Rudnick}}]{DBR2014}
\bibinfo{author}{\bibfnamefont{D.}~\bibnamefont{Dantchev}},
  \bibinfo{author}{\bibfnamefont{J.}~\bibnamefont{Bergknoff}},
  \bibnamefont{and} \bibinfo{author}{\bibfnamefont{J.}~\bibnamefont{Rudnick}},
  \bibinfo{journal}{Phys. Rev. E} \textbf{\bibinfo{volume}{89}},
  \bibinfo{pages}{042116} (\bibinfo{year}{2014}).

\bibitem[{\citenamefont{Diehl et~al.}(2012)\citenamefont{Diehl, Gr\"{u}neberg,
  Hasenbusch, Hucht, Rutkevich, and Schmidt}}]{DGHHRS2012}
\bibinfo{author}{\bibfnamefont{H.~W.} \bibnamefont{Diehl}},
  \bibinfo{author}{\bibfnamefont{D.}~\bibnamefont{Gr\"{u}neberg}},
  \bibinfo{author}{\bibfnamefont{M.}~\bibnamefont{Hasenbusch}},
  \bibinfo{author}{\bibfnamefont{A.}~\bibnamefont{Hucht}},
  \bibinfo{author}{\bibfnamefont{S.~B.} \bibnamefont{Rutkevich}},
  \bibnamefont{and} \bibinfo{author}{\bibfnamefont{F.~M.}
  \bibnamefont{Schmidt}}, \bibinfo{journal}{EPL}
  \textbf{\bibinfo{volume}{100}}, \bibinfo{pages}{10004}
  (\bibinfo{year}{2012}).

\bibitem[{\citenamefont{{Diehl}
  et~al.}(2014{\natexlab{b}})\citenamefont{{Diehl}, {Gr{\"u}neberg},
  {Hasenbusch}, {Hucht}, {Rutkevich}, and {Schmidt}}}]{DGHHRS2014}
\bibinfo{author}{\bibfnamefont{H.~W.} \bibnamefont{{Diehl}}},
  \bibinfo{author}{\bibfnamefont{D.}~\bibnamefont{{Gr{\"u}neberg}}},
  \bibinfo{author}{\bibfnamefont{M.}~\bibnamefont{{Hasenbusch}}},
  \bibinfo{author}{\bibfnamefont{A.}~\bibnamefont{{Hucht}}},
  \bibinfo{author}{\bibfnamefont{S.~B.} \bibnamefont{{Rutkevich}}},
  \bibnamefont{and} \bibinfo{author}{\bibfnamefont{F.~M.}
  \bibnamefont{{Schmidt}}}, \bibinfo{journal}{Phys. Rev. E} \textbf{\bibinfo{volume}{89}}, \bibinfo{pages}{062123}
  (\bibinfo{year}{2014}{\natexlab{b}}).

\bibitem[{\citenamefont{Barber et~al.}(1974)\citenamefont{Barber, Jasnow,
  Singh, and Weiner}}]{BJSW74}
\bibinfo{author}{\bibfnamefont{M.~N.} \bibnamefont{Barber}},
  \bibinfo{author}{\bibfnamefont{D.}~\bibnamefont{Jasnow}},
  \bibinfo{author}{\bibfnamefont{S.}~\bibnamefont{Singh}}, \bibnamefont{and}
  \bibinfo{author}{\bibfnamefont{R.~A.} \bibnamefont{Weiner}},
  \bibinfo{journal}{Journal of Physics C: Solid State Physics}
  \textbf{\bibinfo{volume}{7}}, \bibinfo{pages}{3491} (\bibinfo{year}{1974}).

\bibitem[{\citenamefont{Bray and Moore}(1977)}]{BM77}
\bibinfo{author}{\bibfnamefont{A.~J.} \bibnamefont{Bray}} \bibnamefont{and}
  \bibinfo{author}{\bibfnamefont{M.~A.} \bibnamefont{Moore}},
  \bibinfo{journal}{J. Phys. A: Math. Gen.} \textbf{\bibinfo{volume}{10}},
  \bibinfo{pages}{1927} (\bibinfo{year}{1977}).

\bibitem[{\citenamefont{{Diehl} and {Rutkevich}}(2014)}]{DR2014}
\bibinfo{author}{\bibfnamefont{H.~W.} \bibnamefont{{Diehl}}} \bibnamefont{and}
  \bibinfo{author}{\bibfnamefont{S.~B.} \bibnamefont{{Rutkevich}}},
  \bibinfo{journal}{J. Phys. A: Math. Theor.} \textbf{\bibinfo{volume}{47}},
  \bibinfo{pages}{145004} (\bibinfo{year}{2014}), \eprint{1401.1357}.

\bibitem[{\citenamefont{Dantchev}(1996)}]{D96}
\bibinfo{author}{\bibfnamefont{D.}~\bibnamefont{Dantchev}},
  \bibinfo{journal}{Phys. Rev. E} \textbf{\bibinfo{volume}{53}},
  \bibinfo{pages}{2104} (\bibinfo{year}{1996}).

\bibitem[{\citenamefont{Dantchev}(1998)}]{D98}
\bibinfo{author}{\bibfnamefont{D.~M.} \bibnamefont{Dantchev}},
  \bibinfo{journal}{Phys. Rev. E} \textbf{\bibinfo{volume}{58}},
  \bibinfo{pages}{1455} (\bibinfo{year}{1998}).

\bibitem[{\citenamefont{Dantchev and Gr\"{u}neberg}(2009)}]{DG2009}
\bibinfo{author}{\bibfnamefont{D.}~\bibnamefont{Dantchev}} \bibnamefont{and}
  \bibinfo{author}{\bibfnamefont{D.}~\bibnamefont{Gr\"{u}neberg}},
  \bibinfo{journal}{Phys. Rev. E} \textbf{\bibinfo{volume}{79}},
  \bibinfo{eid}{041103} (pages~\bibinfo{numpages}{16}) (\bibinfo{year}{2009}).

\bibitem[{\citenamefont{Shapiro and Rudnick}(1986)}]{SR86}
\bibinfo{author}{\bibfnamefont{J.}~\bibnamefont{Shapiro}} \bibnamefont{and}
  \bibinfo{author}{\bibfnamefont{J.}~\bibnamefont{Rudnick}},
  \bibinfo{journal}{J. Stat. Phys.} \textbf{\bibinfo{volume}{43}},
  \bibinfo{pages}{51} (\bibinfo{year}{1986}).

\bibitem[{\citenamefont{Fisher and Barber}(1972)}]{FB72}
\bibinfo{author}{\bibfnamefont{M.~E.} \bibnamefont{Fisher}} \bibnamefont{and}
  \bibinfo{author}{\bibfnamefont{M.~N.} \bibnamefont{Barber}},
  \bibinfo{journal}{Phys. Rev. Lett.} \textbf{\bibinfo{volume}{28}},
  \bibinfo{pages}{1516} (\bibinfo{year}{1972}).

\bibitem[{\citenamefont{Barber}(1983)}]{Ba83}
\bibinfo{author}{\bibfnamefont{M.~N.} \bibnamefont{Barber}}, in
  \bibinfo{booktitle}{{\it Phase Transitions and Critical Phenomena}}, edited
  by \bibinfo{editor}{\bibfnamefont{C.}~\bibnamefont{Domb}} \bibnamefont{and}
  \bibinfo{editor}{\bibfnamefont{J.~L.} \bibnamefont{Lebowitz}}
  (\bibinfo{publisher}{Academic, London}, \bibinfo{year}{1983}),
  vol.~\bibinfo{volume}{8}, p. \bibinfo{pages}{145}.

\bibitem[{\citenamefont{Privman}(1990)}]{P90}
\bibinfo{author}{\bibfnamefont{V.}~\bibnamefont{Privman}},
  \bibinfo{title}{{\it Finite Size Scaling and Numerical Simulations of
  Statistical Systems}} (\bibinfo{publisher}{World Scientific, Singapore},
  \bibinfo{year}{1990}), chap. \bibinfo{chapter}{Finite-size scaling theory},
  p.~\bibinfo{pages}{1}.

\bibitem[{\citenamefont{Brankov et~al.}(2000)\citenamefont{Brankov, Dantchev,
  and Tonchev}}]{BDT2000}
\bibinfo{author}{\bibfnamefont{J.~G.} \bibnamefont{Brankov}},
  \bibinfo{author}{\bibfnamefont{D.~M.} \bibnamefont{Dantchev}},
  \bibnamefont{and} \bibinfo{author}{\bibfnamefont{N.~S.}
  \bibnamefont{Tonchev}}, \bibinfo{title}{{\it The Theory of Critical
  Phenomena in Finite-Size Systems - Scaling and Quantum Effects}}
  (\bibinfo{publisher}{World Scientific, Singapore}, \bibinfo{year}{2000}).

\end{thebibliography}
\end{document}